\documentclass[twocolumn,prb,superscriptaddress,preprintnumbers,amsmath,amssymb,floatfix]{revtex4}
\usepackage{graphicx}
\usepackage[dvips]{color}
\usepackage{dcolumn}
\usepackage{bm}
\usepackage{color}
\usepackage{appendix}
\begin{document}
\newcommand{\Co}{BaCoO$_{3}$}

\title{Spin-orbit coupling and crystal-field distortions for a low-spin $3d^5$ state in BaCoO$_{3}$ }

\author{Y.~Y.~Chin}
 \affiliation{Department of Physics, National Chung Cheng University, 168, Sec. 1, University Rd., Min-Hsiung, Chiayi 62102, Taiwan}
 \affiliation{National Synchrotron Radiation Research Center, 101 Hsin-Ann Road, Hsinchu 30076, Taiwan}
 \author{Z.~Hu}
 \affiliation{Max Planck Institute for Chemical Physics of Solids, N{\"o}thnitzer Stra{\ss}e 40, 01187 Dresden, Germany}
\author{H.-J.~Lin}
 \affiliation{National Synchrotron Radiation Research Center, 101 Hsin-Ann Road, Hsinchu 30076, Taiwan}
\author{S.~Agrestini}
 \affiliation{Max Planck Institute for Chemical Physics of Solids, N{\"o}thnitzer Stra{\ss}e 40, 01187 Dresden, Germany}
\author{J.~Weinen}
 \affiliation{Max Planck Institute for Chemical Physics of Solids, N{\"o}thnitzer Stra{\ss}e 40, 01187 Dresden, Germany}
\author{C.~Martin}
 \affiliation{Laboratoire CRISMAT, UMR 6508 CNRS-ENSICAEN, 6 bd Mar\'echal Juin, 14050 Caen Cedex, France}
\author{S.~H\'ebert}
 \affiliation{Laboratoire CRISMAT, UMR 6508 CNRS-ENSICAEN, 6 bd Mar\'echal Juin, 14050 Caen Cedex, France}
 \author{A.~Maignan}
 \affiliation{Laboratoire CRISMAT, UMR 6508 CNRS-ENSICAEN, 6 bd Mar\'echal Juin, 14050 Caen Cedex, France}
\author{A.~Tanaka}
 \affiliation{Department of Quantum Matter, ADSM, Hiroshima University, Higashi-Hiroshima 739-8530, Japan }
\author{J.~C.~Cezar}
 \affiliation{European Synchrotron Radiation Facility, 6 Rue Jules Horowitz, BP 220, 38043, Grenoble, France}
\author{N.~B.~Brookes}
 \affiliation{European Synchrotron Radiation Facility, 6 Rue Jules Horowitz, BP 220, 38043, Grenoble, France}
\author{Y.-F.~Liao}
\affiliation{National Synchrotron Radiation Research Center, 101 Hsin-Ann Road, Hsinchu 30076, Taiwan}
\author{K.-D.~Tsuei}
\affiliation{National Synchrotron Radiation Research Center, 101 Hsin-Ann Road, Hsinchu 30076, Taiwan}
 \author{C.~T.~Chen}
 \affiliation{National Synchrotron Radiation Research Center, 101 Hsin-Ann Road, Hsinchu 30076, Taiwan}
\author{D.~I.~Khomskii}
 \affiliation{Institute of Physics II, University of Cologne, Z{\"u}lpicher Stra{\ss}e 77, 50937 Cologne, Germany}
\author{L.~H.~Tjeng}
 \affiliation{Max Planck Institute for Chemical Physics of Solids, N{\"o}thnitzer Stra{\ss}e 40, 01187 Dresden, Germany}

\begin{abstract}
We have studied the electronic structure of BaCoO$_3$ using soft x-ray absorption spectroscopy at the Co-$L_{2,3}$ and O-$K$ edges, magnetic circular dichroism at the Co-$L_{2,3}$ edges, as well as valence band hard x-ray photoelectron spectroscopy. The quantitative analysis of the spectra established that the Co ions are in the formal low-spin tetravalent 3$d^5$ state and that the system is a negative charge transfer Mott insulator. The spin-orbit coupling plays also an important role for the magnetism of the system. At the same time, a trigonal crystal field is present with sufficient strength to bring the 3$d^5$ ion away from the $J_{eff} = 1/2$ state. The sign of this crystal field is such that the $a_{1g}$ orbital is doubly occupied, explaining the absence of a Peierl's transition in this system which consists of chains of face-sharing CoO$_6$ octahedra.  Moreover, with one hole residing in the $e_g^{\pi}$, the presence of an orbital moment and strong magneto-crystalline anisotropy can be understood. Yet, we also infer that crystal fields with lower symmetry must be present to reproduce the measured orbital moment quantitatively, thereby suggesting the possibility for orbital ordering to occur in BaCoO$_3$.
\end{abstract}

\maketitle

\section{Introduction}

Cobalt oxides have generated considerable attention in the scientific community due to their complex and large diversity of physical phenomena, such as metal-insulator transitions \cite{Raccah67,Martin97,Imada98}, superconductivity \cite{Takada03,Schaak03,Takada04,Ohta11}, large magnetoresistance \cite{Perez98}, high thermoelectric power \cite{Terasaki97, Martin97,Masset00}, and also high catalytic activity for energy storage applications \cite{Suntivich11,Xu16,Liang11,Jeen13}. This richness of electronic and magnetic properties is closely related not only to the possibility of stabilizing cobalt in different valence states but also due to the so-called spin-state degree of freedom
\cite{Goodenough58,Goodenough71,Sugano70,Potze95,Haverkort06,Chen14,Ou16}.
For example, in an octahedral coordination, Co$^{3+}$ or Co$^{4+}$ ions, which have the formal $d^6$ or $d^5$ configuration, respectively, can exist in three possible spin states: a high-spin (HS) state, a low-spin (LS) state, and also even an intermediate-spin (IS) state.

BaCoO$_3$ is a fascinating cobalt oxide with various intriguing aspects. The crystal structure consists of one-dimensional (1D) $c$-axis chains of face-sharing CoO$_6$ octahedra forming a two-dimensional (2D) triangular lattice in the $ab$-plane \cite{Taguchi77,Raghu91,Yamaura99}. BaCoO$_3$ can be considered as belonging to the material class of A$_{n+2}$Co$_{n+1}$O$_{3n+3}$ (A =Ca, Sr, Ba, n $\rightarrow\infty$)
\cite{Sugiyama05,Sugiyama06,Hebert07}. Depending on $n$ and on the A ion, the competition between the 1D and 2D interactions in this material class can generate peculiar transport and magnetic properties such as successive magnetic transitions \cite{Achiwa69,Nozaki07} and magnetization plateaus
\cite{Hardy04,Maignan04,Agrestini08,Fleck10,Agrestini11}, unusually large magnetocrystalline anisotropy \cite{Burnus08} and collinear-magnetism-driven ferroelectricity  \cite{Choi08}, as well as the phenomenon of quantum tunneling of the magnetization \cite{Maignan04}.

The high temperature magnetic susceptibility of BaCoO$_3$ shows an effective magnetic moment of 2.3 $\mu_B$, which was taken as a sign for the existence of LS Co$^{4+}$ \cite{Yamaura99,Wang15}, although it is larger than the spin-only value of 1.73 $\mu_B$ for an $S=1/2$ ion. Neutron powder diffraction (NPD) \cite{Nozaki07} and $\mu$SR experiments  \cite{Sugiyama05,Nozaki07} indicated the presence of a 3D AFM coupling below T$_N$ = 15 K. Between 15 K and 53 K, 2D ferromagnetism takes place and above 53 K there is superparamagnetism followed by the Curie law for temperatures above 250 K \cite{Nozaki07,Sugiyama05,Sugiyama06}. The magnetic Bragg reflections in NPD can be indexed with FM coupling intra-chain and AFM coupling between the chains with a propagation vector $k$ = (1/3,1/3,0) \cite{Nozaki07} and a magnetic moment of 0.53 $\mu_B$, suggesting a geometric frustration for the triangular lattice in the ab-plane.

BaCoO$_3$ is a small-gap semiconductor based on temperature dependent resistivity measurements \cite{Raghu91,Yamaura99,Wang15}. It is actually quite remarkable that it is not metallic. In view of the very high formal oxidation state of 4+, one may expect that the oxygen $2p$ to cobalt $3d$
charge transfer energy is negative, and that the system should show a p-type metallic behavior according to the Zaanen-Sawatzky-Allen phase diagram \cite{Zaanen85}. This apparently does not happen and is thus different from, for example, SrCoO$_3$ \cite{Potze95}, also a octahedral Co$^{4+}$ system, which indeed shows a metallic signature in its resistivity.

Several \emph{ab-initio} band structure calculations have been carried out to explain the physical properties of BaCoO$_3$ \cite{Felser99,Cacheiro03,Pardo04,Pardo05,Pardo06a,Pardo06b,Pardo07}. It was concluded that a Peierls transition in the 1D $c$-axis chain is unlikely to be the reason for the insulating state \cite{Felser99}. Later calculations took into account electron correlation effects at the Co sites in the  so-called LDA+U approach \cite{Pardo04,Pardo05,Pardo06a,Pardo06b,Pardo07}
to reproduce the insulating state and to explain the magnetic structure. These calculations also proposed that the total energy of the system can be lowered if an orbital ordering is allowed to occur within the $c$-axis chain.

In view of the recent frantic search for the materialization of quantum spin liquids and Kitaev model based on Ir and Ru compounds with the LS octahedral $d^5$ configuration \cite{Jackeli09,Chaloupka10,Kitaev06}, it would be useful to know whether the spin-orbit interaction of the Co ion can also stabilize the $J_{eff}$=$1/2$ state in BaCoO$_3$. Interestingly, BaCoO$_3$ was also mentioned explicitly in recent theoretical studies \cite{Kugel15,Khomskii16} as a candidate material for SU(4) physics to occur: a system with a highly symmetric Hamiltonian containing orbital and spin interactions of the Heisenberg type \cite{Kugel73,Kugel82,Frischmuth99} showing, for example, gapless spin and orbital waves.

Our objective here is to determine experimentally, using x-ray absorption spectroscopy (XAS) and x-ray magnetic circular dichroism (XMCD), the local electronic structure of the Co ions in BaCoO$_3$. In particular, we aim to determine the charge, spin, and orbital state of the Co and its relation to the local coordination and crystal structure. We also need to determine
spectroscopically that the material is an insulator in order to provide a justification for the various approximations to model its magnetic properties. With this we hope to shed light on why there is no Peierls distortion in this system, why the measured effective magnetic moment
deviates from the spin-only value, and whether the conditions for a  $J_{eff}$=$1/2$ state, for the orbital ordering or for other orbital physics phenomena as proposed by the theoretical studies mentioned above can be met.

\section{Experiments}

The BaCoO$_3$ ceramics were prepared by a two-step process as described elsewhere \cite{Hebert07}. The Co-$L_{2,3}$ and the O-$K$ XAS spectra were measured in the total electron yield mode (TEY) at the Dragon beam line of the National Synchrotron Radiation Research Center in Taiwan. The measurements were carried out at room temperature. The Co-$L_{2,3}$ XMCD spectra were recorded at the ID8 beam line of the European Synchrotron Radiation Facility (ESRF) in Grenoble under a magnetic field of 5 T at 50 K. Below 50 K, the sample became strongly charging and could not be measured in the TEY mode, indicating insulating properties. CoO and NiO single crystals were measured simultaneously in another chamber for energy calibration.

The hard x-ray photoelectron spectroscopy (HAXPES) experiment has been carried out at the Max-Planck-NSRRC end-station at the Taiwan undulator beamline BL12XU at SPring-8 in Hyogo, Japan. The photon beam was linearly polarized with the electrical field vector in the plane of the storage ring (i.e. horizontal) and the photon energy was set at about 6.5 keV. The experimental set-up has two MB Scientific A-1 HE analyzers. We have used the analyzer which was mounted vertically. The direction of the photoelectrons was thus perpendicular to the electrical field vector and the Poynting vector of the beam. The overall energy resolution was 0.35 eV and the Fermi level was calibrated using polycrystalline gold. The BaCoO$_3$ sample was cleaved $in$ $situ$ in order to have a clean surface. The measurements were done at room temperature.

\section{Results and Discussion}

\subsection{Valence state and orbital occupation}

\begin{figure}
\center
\includegraphics[width=0.49 \textwidth]{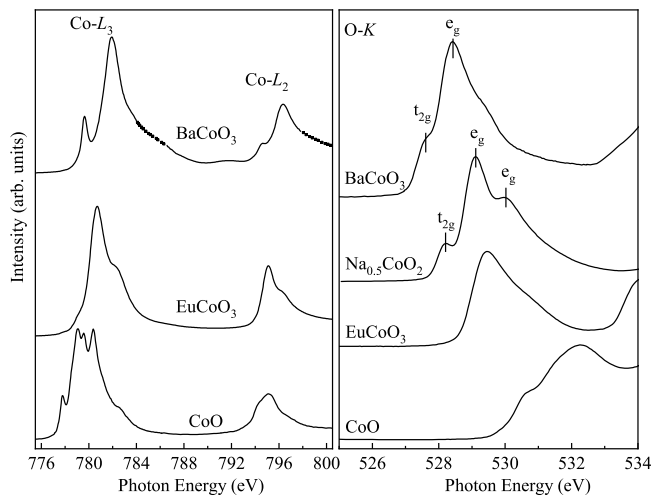}
  \caption{
   Left panel: Experimental Co-$L_{2,3}$ XAS spectra of BaCoO$_3$, EuCoO$_3$ (from \cite{Hu04}) and CoO. Right panel: Experimental O-$K$ XAS spectra of BaCoO$_3$, Na$_{0.5}$CoO$_2$ (from \cite{Lin10}), EuCoO$_3$ (from \cite{Hu04}), and CoO.
   }
\label{BaCo}
\end{figure}

In Fig. \ref{BaCo} (left panel), we present the Co-$L_{2,3}$ XAS spectra of BaCoO$_3$ taken at room temperature together with those of EuCoO$_3$ (from \cite{Hu04}) as a Co$^{3+}$ reference and CoO as a Co$^{2+}$ reference. For the spectrum of BaCoO$_3$, we have removed the Ba-$M_{4,5}$ white lines located at 784 eV and 798 eV using the Ba-$M_{4,5}$ spectrum of BaFeO$_3$ \cite{Hu12}. It is well known that XAS spectra at the 3d transition metal (TM) $L_{2,3}$ edges are highly sensitive to the valence state: an increase of the valence state of the metal ion causes a shift of the XAS $L_{2,3}$ spectra towards higher energies \cite{Mitra03,Burnus08}. In Fig. \ref{BaCo}, we can clearly observe a shift of the center of gravity of the $L_3$ spectrum to higher photon energies from CoO to EuCoO$_3$ and to BaCoO$_3$ by about 1 eV each time. This observation indicates that the formal valence of the Co ions in BaCoO$_3$ is 4+.

By studying the O-$K$ XAS, we can check the valence state, and furthermore, also identify the occupied orbitals of the Co ions in BaCoO$_3$. In Fig. \ref{BaCo} (right panel), the structures from 528 eV to 533 eV are due to transitions from the O 1$s$ core level to the O 2$p$ orbitals which are hybridized with the unoccupied Co 3$d$  $t_{2g}$ and $e_g$ states. From the bottom to the top, one can see a gradual shift of the pre-edge peak in the O-$K$ XAS spectra to lower energies for an increase of the formal valence from Co$^{2+}$ in CoO to Co$^{3+}$ in EuCoO$_3$ \cite{Hu04}, to Co$^{3.5+}$ in Na$_{0.5}$CoO$_2$ \cite{Lin10} and further to Co$^{4+}$ in BaCoO$_3$. This energy lowering of the pre-edge peak reflects the increase of the valence state of transition metal ions in oxides, as known from previous studies \cite{Hu01,Wu05,Mizokawa13}. 

The pre-edge peak structure of the O-$K$ XAS provides also information about the orbital occupations and the possible spin states. The main peak of the EuCoO$_3$ spectrum at 529.5 eV can be assigned to transitions to the fourfold degenerate $e_g$ holes of the LS Co $3d^6$ configuration. The extra peak
appearing at the lower energy of 528.2 eV in Na$_{0.5}$CoO$_2$ \cite{Lin10} and 527.5 eV for BaCoO$_3$ can then be assigned to the presence of an additional hole in the $t_{2g}$ shell. This implies that the Co$^{4+}$ ion has the LS configuration: with one $t_{2g}$ and four $e_g$ holes, the electron orbital occupation is $t_{2g}^5$. The slight energy shift to lower photon energies in the spectrum of BaCoO$_3$ with respect to that of Na$_{0.5}$CoO$_2$ reflects the higher valence state of Co in the former. We would like to note that the O-$K$ XAS spectrum of BaCoO$_3$ is quite  different from that of La$_{1-x}$Sr$_x$CoO$_3$ which has a different orbital occupation and therefore different spin-state (IS) for its Co$^{4+}$ ions \cite{Okamoto00}, confirming our analysis. The orbital occupation and the LS state in BaCoO$_3$ is consistent with the very short Co-O distance of 1.874 ${\AA}$ in it \cite{Taguchi77}.

\subsection{Presence of band gap}

\begin{figure}[h]
\center
 \includegraphics[width=0.45 \textwidth]{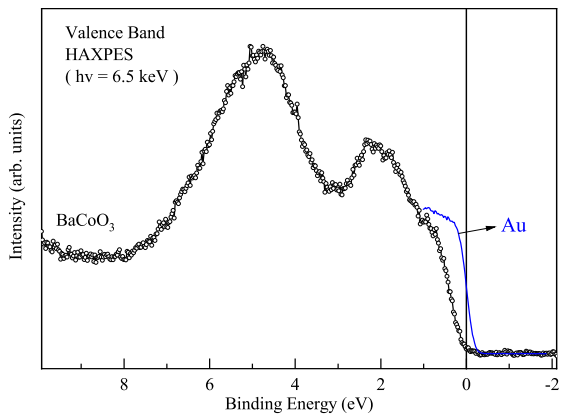}
\vspace{-4mm}
\caption{Valence band HAXPES spectra of BaCoO$_3$ (black circles) and Au (blue curve) for energy calibration.}
\label{VBBa}
\end{figure}

The valence band spectrum of BaCoO$_3$ taken with the bulk-sensitive HAXPES method is displayed in Fig. \ref{VBBa}. The Fermi-edge of Au metal is also measured to serve as reference. The most relevant information that we would like to extract from this room temperature spectrum is that the spectral weight of BaCoO$_3$ at the Fermi level is negligible. This agrees well with the semiconducting behavior as observed in resistivity measurements \cite{Raghu91,Yamaura99,Wang15}. Comparing the leading edge of the BaCoO$_3$ valence band with the Au Fermi-edge, we can estimate that the band gap is about 0.3 eV. Here we assume that the bottom of the conduction band is pinned at the Fermi level. The true band gap value could, of course, be larger if the Fermi level is pinned by in-gap states, but at the moment we have no information about the energy position of the bottom of the conduction
band. Despite these uncertainties, we can safely conclude that BaCoO$_3$ is truly an insulating material with a band gap of several tenths of an eV.

The finding of a band gap is important for the quantitative modeling of the local electronic structure of BaCoO$_3$. We now can meaningfully use a single-site configuration interaction approach which includes the effect of the lattice on the local Co ion by taking an effective CoO$_6$ cluster only. Such a single-site approach may not be valid if the system is a metal with substantial inter-site or inter-cluster charge fluctuations. As mentioned above, a $p$-type metal behavior could be envisioned within the Zaanen-Sawatzky-Allen phase diagram \cite{Zaanen85} for high oxidation state transition metal oxides. Apparently, this does not materialize for BaCoO$_3$, probably due to the fact that the ligand holes have poor inter-site or inter-cluster hopping integrals due to the face-sharing nature of the CoO$_6$
octahedra. Here one can envision that an O 2p orbital which is $\pi$ bonded to the $t_{2g}$ orbital of a particular Co site has poor $\pi$ bonding with the $t_{2g}$ orbital of the neighboring Co due to the fact that the Co-O-Co bond angle of about 78 degrees \cite{Taguchi77,Calle08} is not too far from 90 degrees. The situation for oxides consisting of corner-sharing octahedra, e.g. SrCoO$_3$ \cite{Potze95,Pezdika93,Kunes12} and Sr$_2$CoO$_4$ \cite{Matsuno04,Lee06}, is clearly different.

\subsection{Charge-transfer energy and spin-state}

\begin{figure}[h]
\center
\includegraphics[width=0.45 \textwidth]{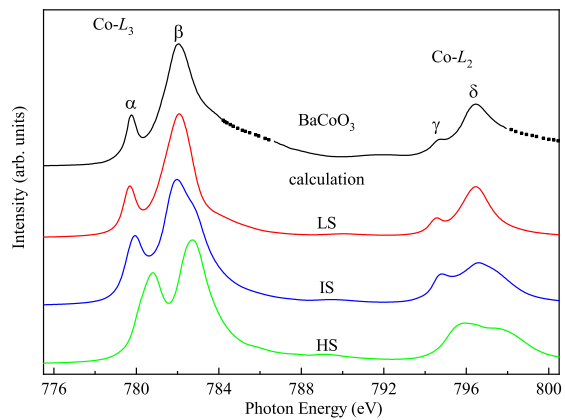}
\caption{
     Calculated Co-$L_{2,3}$ XAS spectra for a CoO$_6$ cluster with the 3$d^5$ low-spin (LS), intermediate-spin (IS) and high-spin (HS) state configurations. The O $2p$ to Co $3d$ charge transfer energy in the calculations is negative: -3.5 eV. Characteristic features of the experimental spectrum are labelled with $\alpha$, $\beta$, $\gamma$, and $\delta$ to facilitate the discussion.}
\label{BaCosim}
\end{figure}

We now analyze quantitatively the Co-$L_{2,3}$ XAS spectrum of BaCoO$_3$ from Fig. \ref{BaCo} (left panel, top curve) using a CoO$_6$ cluster model which includes configuration interaction and full atomic multiplet theory \cite{Groot94,Tanaka94}. It is well known that $L_{2,3}$ XAS is sensitive
not only to the charge and orbital state, but also to the spin-state \cite{Hu04,Haverkort06}. Fig. \ref{BaCosim} depicts the results. Here we have calculated the spectra for a Co$^{4+}$O$_6$ cluster with the 3$d^5$ low-spin (LS), intermediate-spin (IS) and high-spin (HS) state configurations
\cite{parameter}. We can clearly observe that the LS scenario gives by far the best match to the experimental spectrum. The energy positions and intensities of the characteristic features labelled $\alpha$, $\beta$, $\gamma$, and $\delta$ are all well reproduced. This establishes firmly that the Co ion is in the formal $S$=$1/2$ $t_{2g}^5$ type of configuration, fully consistent with the O $K$-edge analysis above. The stabilization of the LS state can be attributed to a large effective crystal or ligand field interaction due to the relatively short Co-O distance of 1.874 {\AA}. This is to be contrasted to, for example, SrCoO$_3$ which has a larger Co-O distance of 1.918 {\AA} and has an IS like ground state \cite{Pezdika93,Potze95}.

Important to achieve a good agreement between theory and experiment is to use a negative value for the O $2p$ to Co $3d$ charge transfer energy in the calculation: we have taken -3.5 eV. This means that the main charge configuration of the Co is not 3$d^5$ (8\% weight) but mainly
3$d^6\underline{L}$ (45\%) and 3$d^7\underline{L}^2$ (40\%) with even also some 3$d^8\underline{L}^3$ (7\%). Here $\underline{L}$ denotes a hole in the O $2p$ ligands. The formal valence and the local symmetry of the Co system is nevertheless still that of a Co$^{4+}$.

In the next sections we will explain further important details that we were able to extract from the calculations, in particular about the effect of the spin-orbit coupling (SOC) and low-symmetry crystal field interactions determining the magnetic properties as well as the absence of a Peierls distortion.

\subsection{Spin-orbit interaction in a $d^5$ system}

\begin{figure}[h]
\center
\includegraphics[width=0.45 \textwidth]{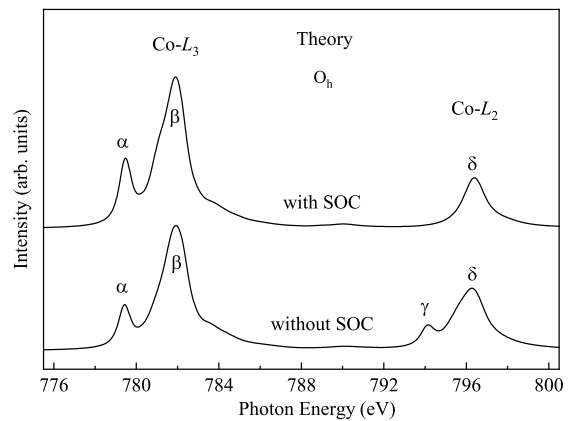}
\caption{
     Calculated Co-$L_{2,3}$ XAS spectra for a CoO$_6$ cluster with the 3$d^5$ low-spin (LS) configuration in $O_h$ symmetry, with and without spin-orbit coupling (SOC).
     }
\label{SOCd5}
\end{figure}

We first assume a pure $O_h$ coordination for the CoO$_6$ cluster and calculate the Co-$L_{2,3}$ XAS spectrum in the LS state scenario. We have done this calculation with the SOC constant set at the atomic value of $\zeta_{SOC}$ = 65 meV and also with the SOC constant set to zero. The result
is shown in Fig. \ref{SOCd5}. We can see that features $\alpha$, $\beta$, and $\delta$ are present for both cases, but we can also see a large difference with respect to feature $\gamma$. Without the SOC, the $L_{2}$ white line contains the pre-peak structure $\gamma$, while such a pre-peak is completely
absent when the SOC is included. Interestingly, the experimental $L_{2}$ white line of BaCoO$_3$ does contain such a pre-peak, although not as strong as in the scenario where the SOC is switched off. 

We now concentrate on the cause of the presence or absence of pre-peak $\gamma$. The LS ground state of a $d^5$ ion in $O_h$ symmetry is given by the $t_{2g}^5$ orbital configuration. The presence of SOC convert this into an effective $J_{eff}$=$1/2$ state, a ground state that one may hope to find in Ir$^{4+}$ and Ru$^{3+}$ compounds \cite{Jackeli09,Chaloupka10}. It was already noticed early on for LS $O_h$-coordinated $4d^5$ transition metal compounds that the $L_2$ edge does not have a pre-peak $\gamma$ \cite{Sham83,Groot94}. The explanation is that the matrix element is zero for the $2p$ to $t_{2g}$ transition at the $L_2$ edge if the ground state is $J_{eff}$=$1/2$ \cite{Sham83,Hu00}. Switching off the SOC, and thus abandoning the $J_{eff}$=$1/2$, allows this matrix element to become  non-zero, and thus pre-peak $\gamma$ to appear, as we can see in Fig. \ref{SOCd5}.

We thus can conclude that the presence of a pre-peak $\gamma$ in the  experimental $L_{2,3}$ XAS spectrum implies that BaCoO$_3$ is not a $J_{eff}$=$1/2$ system. This in turn means that the effect of the SOC must be superseded by a strong crystal field with a symmetry lower than $O_h$.

\subsection{Crystal fields (I): $D_{3d}$}

\begin{figure}
\center
 \includegraphics[width=0.49 \textwidth]{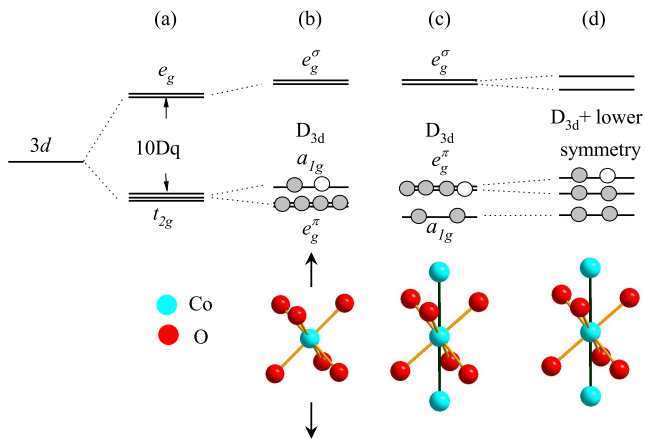}
 \caption{
     Schematic one-electron energy level diagram of a CoO$_6$ cluster in BaCoO$_3$. In $O_h$ coordination (a) the levels are given by the $t_{2g}$ and $e_g$ subshells split by an energy $10Dq$. In the presence of a $D_{3d}$ trigonal distortion, the levels are labeled as $e_g^{\pi}$, $a_{1g}$, and $e_g^{\sigma}$, whereby for a positive
     trigonal crystal field splitting (b) the LS $d^5$  configuration is characterized by a fully occupied $e_g^{\pi}$ and one hole in the $a_{1g}$, and for a negative splitting (c) the LS $d^5$ configuration is described by a fully occupied $a_{1g}$ and one hole in the      $e_g^{\pi}$.  Adding a local lower symmetry distortion (d) split the levels further into five doublets. The influence of the spin-orbit coupling is not included in the diagram.
     }
\label{diagram}
\end{figure}

In lowering the local symmetry of the Co ion from a pure $O_h$, we will first consider a $D_{3d}$ trigonal distortion as this is suggested most naturally from the crystal structure. As depicted in Fig. \ref{diagram}, the one-electron crystal or ligand field energy levels carry the $e_g^{\pi}$,
$a_{1g}$, and $e_g^{\sigma}$ labels in $D_{3d}$ instead of $t_{2g}$ and $e_g$ in $O_h$. Here we are showing two scenarios: one in which the $a_{1g}$ orbital is higher in energy (more unstable) than the $e_g^{\pi}$, to be associated with a positive trigonal crystal field splitting, and one where the $e_g^{\pi}$ is higher and thus a negative splitting. For a LS $d^5$ configuration, a positive splitting would result in a fully occupied $e_g^{\pi}$ and one hole in the $a_{1g}$, while a negative splitting would stabilize a fully occupied $a_{1g}$ and one hole in the $e_g^{\pi}$. The influence of the spin-orbit coupling is not included in the diagram.

\begin{figure}
\center
 \includegraphics[width=0.45 \textwidth]{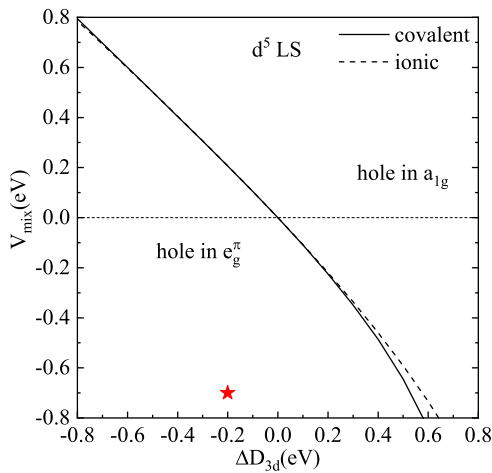}
 \caption{
     Phase diagram for a low-spin Co$^{4+}$ $t_{2g}^5$ configuration in $D_{3d}$ symmetry, displaying the character of the hole occupation as a function of the crystal field parameter $\triangle D_{3d}$ and the mixing parameter $V_{mix}$. The phase diagram has been calculated with the spin-orbit coupling set to zero, and for a covalent (solid line as boundary) and an ionic (dashed line) scenario, see text for further details. The red star in the phase diagram indicates the values for $\triangle D_{3d}$ and $V_{mix}$ needed to reproduce the experimental XAS and XMCD spectra of BaCoO$_3$.
     }
\label{phasediagram}
\end{figure}

The crystal field splitting in $D_{3d}$ symmetry is determined by the crystal field parameter $\triangle D_{3d}$ and the mixing parameter $V_{mix}$, which describes the mixing between the $t_{2g}^{\pm}$ and the $e_g^{\pm}$ leading to the formation of the $e_g^{\pi}$ and $e_g^{\sigma}$ orbitals. Here  $t_{2g}^+$ = $\sqrt{2/3}d_{x^2-y^2}-\sqrt{1/3}d_{xz}$, $t_{2g}^-$ = $\sqrt{2/3}d_{xy}+\sqrt{1/3}d_{yz}$, $e_g^+$ = $\sqrt{1/3}d_{x^2-y^2}+\sqrt{2/3}d_{xz}$, and $e_g^-$ = $\sqrt{1/3}d_{xy}-\sqrt{2/3}d_{yz}$ \cite{Lin10,Haverkort_thesis}. We have calculated the sign of the splitting as a function of these two parameters and display the result in Fig. \ref{phasediagram}. We can clearly see that for negative $\triangle D_{3d}$ and negative $V_{mix}$ the hole will be in the $e_g^{\pi}$ while for positive $\triangle D_{3d}$ and positive $V_{mix}$ it will be in the $a_{1g}$. The calculations have been done with the SOC set to zero. The solid line represents the border between $e_g^{\pi}$ and $a_{1g}$ hole situations in a calculation taking into account the covalency and negative charge transfer energy, while the dashed line is the border in an ionic calculation where the LS state is artificially stabilized by taking
a sufficiently large $O_h$ crystal field splitting.

In trying to determine the values of $\triangle D_{3d}$ and $V_{mix}$ for BaCoO$_3$ from the experimental XMCD spectra, we have carried out the cluster calculations where we have switched on the SOC, i.e. set to its atomic value. However, it turned out that we were not able to find a satisfactory match between the experimental spectra and the simulations within the $D_{3d}$ symmetry. A representative set of the 'not-satisfactory' simulations can be found in the Appendix. We found out that we need to lower the local symmetry of the Co ion even further.

\subsection{Crystal fields (II): lower symmetry}

\begin{figure}[h]
\center
\includegraphics[width=0.45 \textwidth]{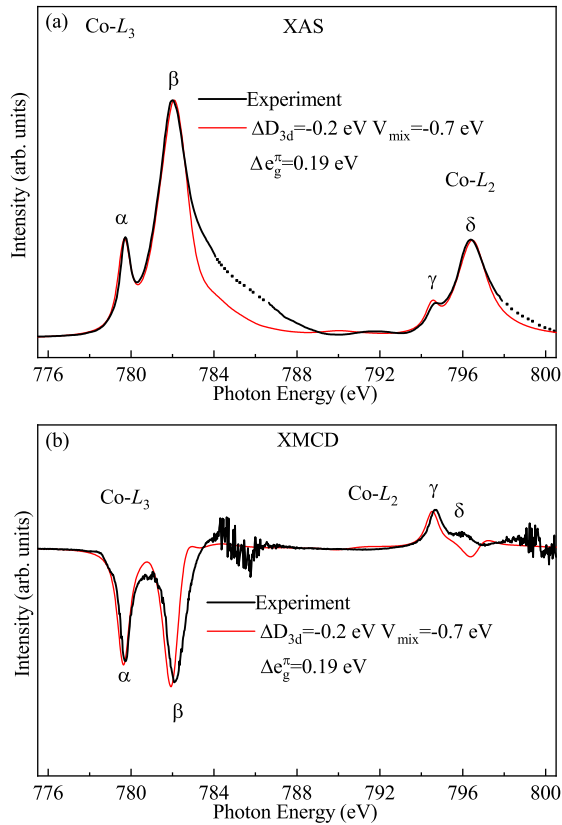}
\caption{
     Experimental Co $L_{2,3}$ (a) XAS and (b) XMCD spectra of BaCoO$_3$ (black lines/dots) together with simulations (red lines) using a CoO$_6$ cluster with the 3$d^5$ low-spin (LS) configuration which includes a trigonal crystal field and lower symmetry crystal field, see text.
     }
\label{BaCoXASXMCD}
\end{figure}

Fig. \ref{BaCoXASXMCD}(a) shows the overlay of the experimental XAS spectrum of BaCoO$_3$ and the simulation which includes a low crystal field that splits the $e_g^{\pi}$ level by about 0.19 eV. The SOC is here also set to its atomic value. We now can observe a very good agreement between the simulation and experiment. The $D_{3d}$ parameters that we used are $\triangle D_{3d}$ = -0.2 eV and $V_{mix}$ = -0.7 eV. We are thus deep in the phase where the hole is the $e_g^{\pi}$ according to Fig. \ref{phasediagram}. This result is somewhat surprising since the trigonal distortion corresponds to the elongation of the octahedron along the $c$-axis, so one may actually expect to have a positive trigonal crystal field splitting. Apparently longer range interactions, due to e.g. the presence of highly charge positive (4+)
nearest neighbor Co ions in the c-axis chain, have a greater influence and make in the end the trigonal crystal field to be effectively negative.

Looking in detail at the $L_{2}$ white line, we can observe that the pre-peak $\gamma$ can be reproduced in the simulation. So unlike in a pure $O_h$ symmetry, the trigonal crystal field causes a mixing between the  $J_{eff}$=$1/2$ and $J_{eff}$=$3/2$ states and this mixing is strong enough that the pre-peak $\gamma$ indeed can show up in the spectrum. This in turn
reiterates that BaCoO$_3$ is not a $J_{eff}$=$1/2$ system.

Fig. \ref{BaCoXASXMCD}(b) displays the XMCD spectrum taken at 50 K with a 5 T magnetic field (black lines/dots). The XMCD signal is pronounced, and it is remarkable that the intensity at the $L_{3}$ white line is heavily negative while that at the $L_{2}$ is only moderately positive. Using the XMCD sum rules developed by Thole, Carra \emph{et al.} \cite{Thole92,Carra93}, we can
directly infer that the negative value for the integrated XMCD intensity is indicative of the presence of an appreciable orbital contribution to the Co magnetic moment. This directly explains why the high temperature magnetic susceptibility of BaCoO$_3$ shows an effective magnetic moment of
2.3 $\mu_B$ \cite{Yamaura99,Wang15} that is larger than the spin-only value of 1.73 $\mu_B$ for an $S=1/2$ ion.

Fig. \ref{BaCoXASXMCD}(b) also shows the simulation for the XMCD using the same parameter set as for the XAS in Fig. \ref{BaCoXASXMCD}(a). The agreement between simulation and experiment is also very good for the XMCD. We  therefore can safely conclude that we have found the proper set of parameter values for BaCoO$_3$.

We would like to note that our finding for the presence of the hole in the $e_g^{\pi}$ orbital is in direct agreement with the orbital moment found from the XMCD since the opposite scenario in which the hole is in the $a_{1g}$ will, in an ionic picture, not produce an orbital moment (we remark that due to the negative charge transfer energy and thus the presence of the $d^6\underline{L}$ some amount of orbital moment will be present, ca. 0.2-0.3 $\mu_B$).

We also need to point out that the presence of a low symmetry crystal field that splits the $e_g^{\pi}$ level is necessary to explain the XMCD orbital moment quantitatively. Without such a splitting, a hole in the degenerate $e_g^{\pi}$ orbital would carry an orbital moment of 1.04 $\mu_B$. The simulation in Fig. \ref{BaCoXASXMCD}(b) reveals that the orbital moment from the XMCD is about 0.52 $\mu_B$. Thus the low symmetry crystal field is necessary to moderate the effect of the SOC.

\subsection{Total energy diagram}

\begin{figure}
\center
 \includegraphics[width=0.45 \textwidth]{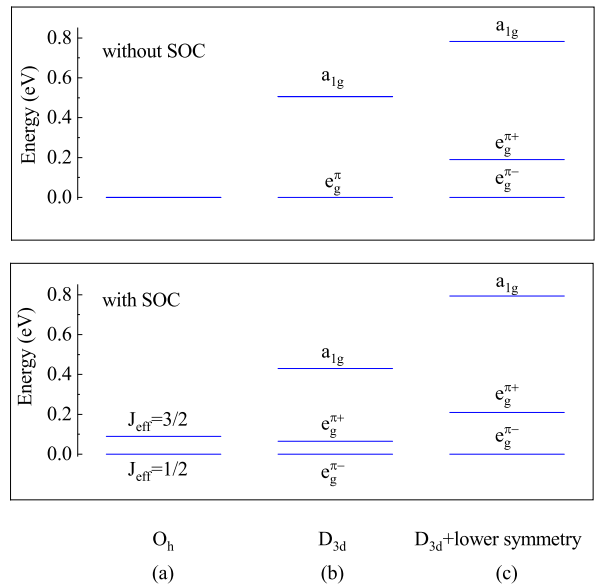}
 \vspace{-3.5cm}
 \caption{
     Total energy level diagram of a $3d^5$ low-spin (LS) CoO$_6$ cluster in BaCoO$_3$. (a) In $O_h$ coordination, (b) with a trigonal distortion, and (c) with additional low symmetry crystal field. The diagram is playing the situation with and without the spin-orbit coupling. Only the lowest three states are shown.
     }
\label{total_energy_diagram}
\end{figure}

Fig. \ref{total_energy_diagram} shows the total energy level diagram for the LS $3d^5$ CoO$_6$ cluster, and summarizes the scenarios that we have investigated. Starting with a pure $O_h$ coordination (a), the LS Co$^{4+}$ ion will have its five electrons in the $t_{2g}$ shell. The SOC will split this six-fold degenerate $^2T_2$ state \cite{Sugano70} into a two-fold degenerate $J_{eff}$=$1/2$ ground state and a four-fold degenerate $J_{eff}$=$3/2$ first excited state. The presence of the pre-peak $\gamma$ in the experimental $L_2$ XAS white line indicates that the $J_{eff}$=$1/2$ ground state did not materialize in BaCoO$_3$. Inclusion of a trigonal distortion produces a two-fold degenerate $a_{1g}$ and a four-fold degenerate $e_g^{\pi}$ state. The presence of the SOC splits this state further, and the result is that the ground state carries a large orbital moment of about 1.04 $\mu_B$. This is too large in comparison to the value obtained from the XMCD experiments, so that a lower crystal field is needed to split the $e_g^{\pi}$ with an energy separation comparable to the SOC. This is shown in (c). In the presence of the SOC, this state produces then a magnetic moment with a more moderate orbital contribution of about 0.52 $\mu_B$.

\subsection{Discussion}

Our finding of a negative trigonal crystal field and thus a hole in the  $e_g^{\pi}$ orbital means that the $a_{1g}$ orbital is doubly occupied. This has important consequences for the discussion about the presence or absence of a Peierls distortion in BaCoO$_3$. The classical Peierls distortion occurs for a one-dimensional metallic system if a doubling of a unit cell by forming dimers can lead to a total energy gain, thereby also opening up a band gap. The starting point or necessary condition is a metallic band before the distortion takes place. In our case, we find that the $a_{1g}$ orbital which is directed along the one-dimensional $c$-axis chain is fully occupied, i.e. there is no degeneracy left to make a metallic band with this $a_{1g}$. This implies that there cannot be a Peierls phenomenon occurring in BaCoO$_3$, at least not on the basis of the $a_{1g}$ band. Our results support the findings from the LDA+U band structure calculations, i.e. calculations which include the effect of Coulomb interactions in the Co $3d$ shell \cite{Pardo04,Pardo05,Pardo06a,Pardo06b,Pardo07}. Those calculations also found that the $a_{1g}$ is completely below the Fermi level, i.e. fully occupied. Calculations without taking into account the on-site Coulomb interaction \cite{Felser99,Cacheiro03}, in contrast, produced an $a_{1g}$ band that crosses the Fermi level, i.e. a metallic band. Our experiment therefore also clearly shows that the insulating nature of BaCoO$_3$ is due to Mott physics.

It is quite interesting that we have found the presence of a crystal field with a symmetry lower than $D_{3d}$, namely to split the $e_g^{\pi}$ levels so that the XAS and XMCD can be well reproduced in the simulations. This lifting of the degeneracy of the $e_g^{\pi}$ orbitals would be compatible
with the proposal by Pardo \emph{et al.} of having an in-chain orbital ordering \cite{Pardo04,Pardo05,Pardo06a,Pardo06b,Pardo07}. This aspect deserves further study using high resolution x-ray diffraction measurements on single crystals.

\section{Conclusions}

We have carried out a detailed study to the local electronic structure of BaCoO$_3$ using soft x-ray absorption, magnetic circular dichroism, and hard x-ray photoemission. We established that the Co ions are in the formal low-spin tetravalent 3$d^5$ state and that this oxide is a negative charge transfer energy system. Remarkably, it is also a Mott insulator despite its negative charge transfer energy. Although the low-spin $d^5$ configuration could in principle produce a $J_{eff} = 1/2$ state, we found that this has not materialized due to the presence of a strong trigonal crystal field. The sign of this crystal field is such that the $a_{1g}$ orbital is doubly occupied, explaining the absence of a Peierls distortion in the $c$-axis chains. With one hole residing in the $e_g^{\pi}$ subshell, the spin-orbit interaction becomes active, leading to the presence of an orbital moment which explains why the measured effective magnetic moment is larger than the spin-only value and why the system has a strong magneto-crystalline  anisotropy. Interestingly, we also found that crystal fields with lower symmetry must be present to reproduce the measured orbital moment quantitatively. This then opens the possibility for orbital ordering to occur in BaCoO$_3$ as proposed by LDA+U calculations on BaCoO$_3$  \cite{Pardo04,Pardo05,Pardo06a,Pardo06b,Pardo07}. To what extend the system can be a SU(4) system or could show spiral orbital order \cite{Kugel73,Kugel82,Frischmuth99} requires further detailed research.

\section{Acknowledgement}

We would like to thank Maurits Haverkort for valuable discussions. We acknowledge the support by the Ministry of Science and Technology of the Republic of China through MOST 107-2112-M-194-001-MY3, by the Deutsche Forschungsgemeinschaft through SFB 1143 (project-id 247310070), and by the Max Planck-POSTECH-Hsinchu Center for Complex Phase Materials. The work of D.I.Kh. was funded by the Deutsche Forschungsgemeinschaft, Project number 277146847 - CRC 1238.

\section{Appendix}

\setcounter{figure}{0}
\makeatletter
\renewcommand{\thefigure}{A\@arabic\c@figure}
\makeatother

\begin{figure}[h]
\center
\includegraphics[width=0.5 \textwidth]{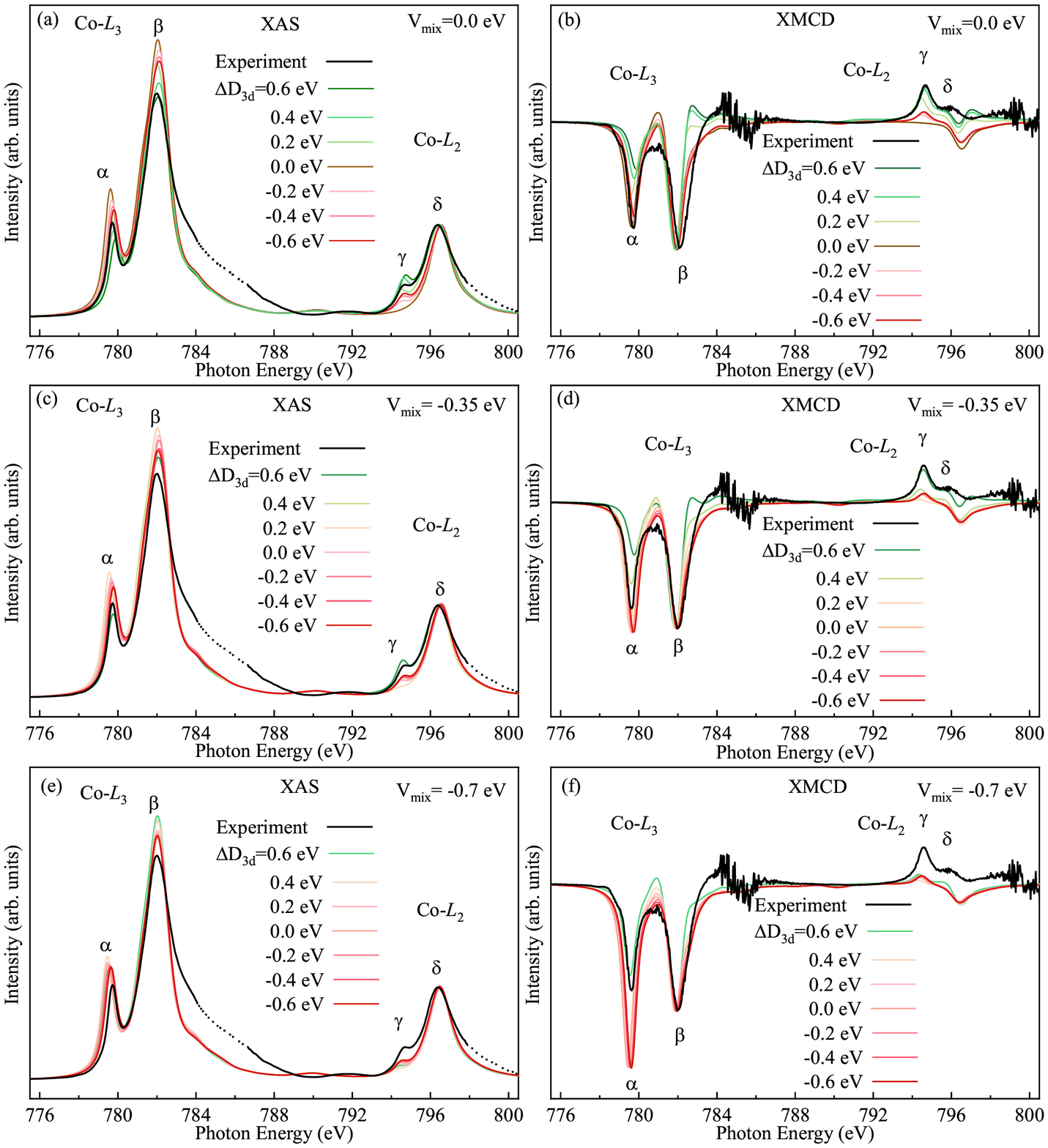}
 \vspace{-3.0cm}
\caption{
     Co $L_{2,3}$ XAS and XMCD spectra of BaCoO$_3$: experiment (black lines/dots) and simulations (colored lines). Panels (a-c-e) show the XAS, and panels (b-d-f) the XMCD data. The simulations were performed for $V_{mix}$ = 0.0 eV (panels a-b), -0.35 eV (panels c-d), and -0.7 eV (panels e-f) and $\triangle D_{3d}$ varying between -0.6 eV and +0.6 eV in steps of 0.2 eV.
     }
\label{A1}
\end{figure}

Fig. \ref{A1} shows the experimental XAS and XMCD spectrum of BaCoO$_3$ together with simulations using a CoO$_6$ cluster in the low-spin configuration. The SOC is set to its atomic value. The local coordination is $D_{3d}$ and the figure illustrates that a satisfactory agreement between experiment and simulation cannot be found for a wide range of parameters $\triangle D_{3d}$ and $V_{mix}$ within this $D_{3d}$ scenario. A lower than $D_{3d}$ symmetry is required
as explained in the main text.

\end{document}